%% file: Lattice2017_230_Sciarra.tex
\documentclass[epj]{webofc}
\usepackage[utf8]{inputenc}
\usepackage[varg]{txfonts}   
\usepackage{booktabs}
\usepackage[svgnames]{xcolor}
\usepackage{graphicx, subfigure}
\usepackage[bookmarks,linktocpage,colorlinks,
            linkcolor = DarkRed,
            urlcolor  = DarkBlue,
            citecolor = DarkGreen]{hyperref}
\usepackage{cleveref}

\wocname{EPJ Web of Conferences}
\woctitle{Lattice2017}
\graphicspath{{Figures/}}

\crefname{figure}{Figure}{Figures}
\crefname{table}{Table}{Tables}
\crefname{enumi}{step}{steps}
\crefname{equation}{Eq.}{Eqs.}
\crefformat{section}{#2\S#1#3}
\Crefformat{section}{#2\S#1#3}

\newcommand{\bahamas}{\texttt{BaHaMAS}}
\newcommand{\betafile}{\mbox{$\beta$-file}}
\newcommand{\clqcd}{\texttt{CL\kern-.15em\textsuperscript{2}QCD}}
\newcommand{\onlinecite}[1]{\nocite{#1}\citenum{#1}} 

\begin{document}

\title{BaHaMAS}
\subtitle{A Bash Handler to Monitor and Administrate Simulations}
\author{%
    \firstname{Alessandro} \lastname{Sciarra}\inst{1}\fnsep
    \thanks{Speaker (indeed substituted by Dr.~Francesca~Cuteri), \href{mailto:sciarra@th.physik.uni-franfurt.de}{sciarra@th.physik.uni-franfurt.de}}
}
\institute{%
    Institut f\"ur Theoretische Physik, J.~W.~Goethe-Universit\"at, 60438 Frankfurt am Main, Germany
}
\abstract{%
    Numerical QCD is often extremely resource demanding and it is not rare to run hundreds of simulations at the same time.
    Each of these can last for days or even months and it typically requires a job-script file as well as an input file with the physical parameters for the application to be run.
    Moreover, some monitoring operations (i.e.\ copying, moving, deleting or modifying files, resume crashed jobs, etc.) are often required to guarantee that the final statistics is correctly accumulated.
    Proceeding manually in handling simulations is probably the most error-prone way and it is deadly uncomfortable and inefficient!
    \bahamas{} was developed and successfully used in the last years as a tool to automatically monitor and administrate simulations.
}
\maketitle

\input{Introduction}
\input{Features}
\input{DesignDecisions}

\input{Examples}

\input{Generalisation}
\input{Conclusions}


\section*{Acknowledgements}

The author would like to thank Owe~Philipsen for the opportunity to develop the software during the stay in Frankfurt, Francesca~Cuteri who constantly advised on the design of the code, Christopher Czaban who contributed to the implementation of some features in the first stage of the project and Aurora Somaglia for having drawn the cover page of the presentation as well as for having provided \bahamas{} with a very nice logo.



\input{Bibliography.bbl}

\end{document}

%% file: Introduction.tex
\section{Introduction}

\begin{figure}
    \centering
    \includegraphics[width=0.85\textwidth]{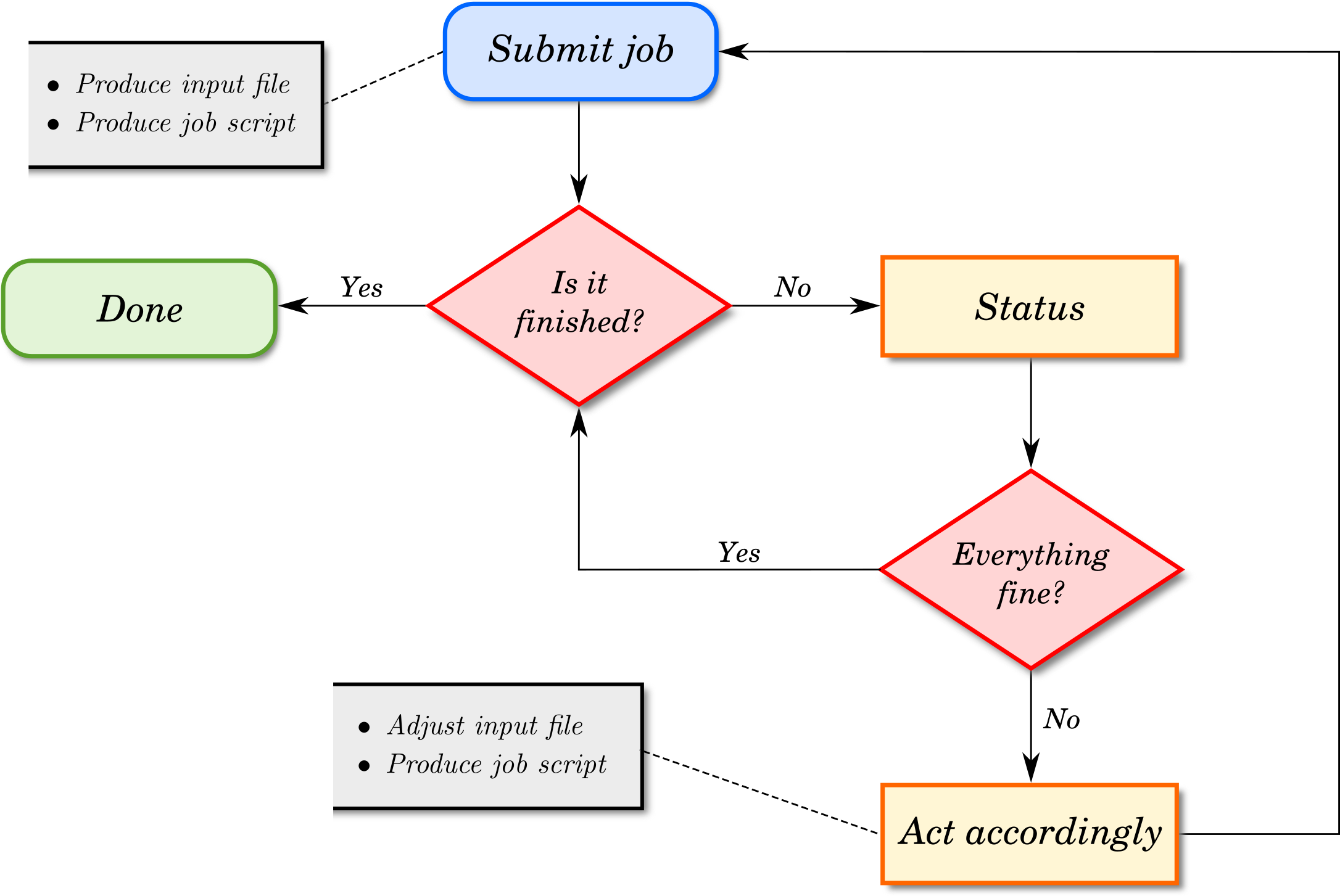}
    \caption{Schematic illustration of the flow of a typical LQCD simulation, which has to be regularly checked.}
    \label{fig:flowSimulation}
\end{figure}

Running a simulation in lattice QCD to produce gauge configurations consists of few, simple steps.
However, because of the high computational cost, it is not enough to just let the simulation start and wait until it is finished.
A routinely work is rather needed and what has to be repeatedly done can be described using a flowchart as done in \cref{fig:flowSimulation}.
Once the simulation has started, it has to be monitored from time to time, since it can take several months before being finished.
Usually, its status is checked on a daily basis.
Whenever a problem occurs, some work has to be done and, consequently, the simulation is continued or resumed from a previous checkpoint.
In the diagram in \cref{fig:flowSimulation}, the routinely work can be easily seen in the inner loop made by the two diamonds and the \emph{Status} step.
In most of the cases, an input file with the physical parameters for the used software as well as a job script with the relevant information for the cluster scheduler are needed to run a simulation.
These have to be produced from scratch at the beginning and then they have to be adjusted or produced again when the simulation has to be stopped and continued.
Moreover, resuming a simulation from a previous checkpoint requires some cleaning procedure.
In principle, all these operations can be carried out when needed just editing, moving or deleting files.
But if one considers that, in a typical real project, hundreds of simulations are run at the same time and are all checked everyday, it should be clear that working by hand is simply not possible.
To provide the reader with some numbers, in~\cite{Cuteri:2015qkq} and~\cite{Philipsen:2016swy} around $1500$ and $1250$ simulations have been run, respectively, and this is also the order of magnitude of runs needed in other ongoing finite temperature lattice QCD studies~\cite{Cuteri:pos17}. 
In $2015$, most functionality of \bahamas{} was not yet implemented and four people were needed to handle simulations in~\cite{Cuteri:2015qkq}.
Already one year later, one single person could administrate a similar number of runs for~\cite{Philipsen:2016swy}.

\medskip

The studies described in references~\onlinecite{Cuteri:2015qkq} and~\onlinecite{Philipsen:2016swy} led to the development of \bahamas{} and widely influenced its design and implementation.
Even though at the beginning it was more like a collection of scripts, it became soon clear that patching here and there~---~a habit to get quickly to a working tool, but which also often leads to a \href{https://en.wikipedia.org/wiki/Big_ball_of_mud}{big ball of mud}~---~would not have led to a \emph{clean code}~\cite{Martin:Clean}. 
Thus, we decided to try to develop more systematically the software, aiming at a readable, maintainable, hard to break, easy to generalise and well tested code.
Needless to say, a software is constantly changing and it will not (probably at any stage) be perfect.
However, we keep working to increase its quality and not only to add new features.

\medskip

\bahamas{} offers an automatised implementation of each step of~\cref{fig:flowSimulation}, together with some useful features to keep under control the full project.
It is publicly available~\cite{bahamas} and it can be easily downloaded either cloning its \texttt{git} repository or as an archive file.
Technical information about the software can be found online in the \href{https://github.com/AG-Philipsen/BaHaMAS/wiki}{Wiki pages} of the repository.
Here in the following, instead, we will begin in \cref{sec:features} focusing more on the main features of the code as well as in \cref{sec:design} on some general design decisions, which are important to know in order to use the software.
Before discussing in \cref{sec:generalisation} the problem of how to provide a software not depending on a particular job scheduler or on a given production code, examples of some functionality of \bahamas{} are presented in \cref{sec:examples}.
Conclusions and future plans follow in \cref{sec:conclusions}.

%% file: Features.tex
\section{The main features of the code}\label{sec:features}

Being written in bash, \bahamas{} does not need to be compiled or installed.
Once cloned its \texttt{git} repository, it can be run straight away.
Nevertheless, to be able to properly work, it needs to be configured with some information, which can be provided in a very intuitive interactive setup procedure.

\bahamas{} provides functionality to automatically handle most of the required steps to produce gauge configurations.
Here, \emph{automatically} means that the user just needs to run a script with some command line option and, not so often, to answer yes-no questions interactively.
More in detail, according to the standard pattern followed in lattice QCD, it is possible to thermalise configurations for given sets of parameters (e.g.\ from a random gauge configuration) and then submit jobs for the production of configurations (with one or several replicas).
Easily, a simulation can be continued from the last checkpoint in case the accumulated statistics is not sufficient and/or it may be resumed from a previous checkpoint, if a problem occurred.

At any time, \bahamas{} can produce an overview of the queued jobs as well as a report on the status of the simulations for a given set of parameters.
To monitor larger projects, depending on the amount of statistics and on the file system of the supercomputer, may require some time and a faster alternative could be needed.
Using the database functionality solves this problem and gives full control of all simulations at once.
From the technical point of view, roughly speaking, the idea is to store the status of the simulations in a file, which is then parsed and filtered on demand to present part of it to the user.
The database can be updated with a given frequency or once per day, typically before checking the status of the runs.

As additional tools next to the main features, an acceptance rate analysis and an automatic cleaning procedure of the output files are implemented.
The former produces a table calculating the acceptance rate of the Monte Carlo run using chunks of history as big as the user desires; the latter is used to eliminate repeated trajectories from the observables files, which usually occur when continuing a simulation from a checkpoint previous to the last produced trajectory. 

Even if handling simulations to produce gauge configurations is its main task, \bahamas{} can also be used to run a code on previously produced configurations.
This is still a young feature, which nevertheless was already successfully used and which shows the potential of the software. 

%% file: DesignDecisions.tex
\section{Some particular design decisions}\label{sec:design}

As any software, \bahamas{} has been designed for precise use cases.
As a consequence, some features serve particular purposes and they could seem limiting for a wider use of the software.
This is certainly true, but it is as well a natural aspect of any young code.
As further discussed in~\cref{sec:generalisation}, it is planned to give more versatility to the software, from several points of view.
Because of that, we decided to discuss here the main decisions made designing the code, without going too much into details that could change in the future.
The reader can refer to the online documentation for further information. 

\subsection*{The physical parameters in the path}

In a typical finite temperature lattice QCD project, many physical parameters have to be varied and many simulations have to be run.
These only differ in few aspects, like the values of the entries in the input file~---~i.e.\ (purely imaginary) chemical potential, quark mass(es), $\beta$, lattice size, etc.
In order to deal effectively with the problem of changing the values of the physical parameters in the input file of each simulation, but also to keep the project folder tidied up as much as possible, \bahamas{} works assuming that a precise folder structure is imposed by the user.
In particular, $5$ parameters must be contained in the path and their values, which are extracted and used by \bahamas{}, need to appear in a given order and, at the moment, to match a fixed format.
A typical path reads something like \texttt{/pathToSubmitDisk/nameOfTheProject/NfX/muiX/massXXXX/ntXX/nsXX}, where the \texttt{X}'s replace actual values.
Since, so far, \bahamas{} has been used to make studies with only degenerate fermions, at zero density~\cite{Cuteri:pos17} or in the Roberge-Weiss plane~\cite{Cuteri:2015qkq,Philipsen:2016swy}, with either Wilson or staggered fermions, there has never been the need of staying maximally general in the path structure.
However, all the information in it is not hard-coded, but it is already contained in global constant variables defined all together in the same file.
This will help in the future to make further generalisations.

\subsection*{The $\beta$-values and the \betafile}

Once fixed the parameters that must be contained in the path~---~i.e.\ when being inside a \texttt{[...]/NfX/muiX/massXXXX/ntXX/nsXX} folder~---~the main remaining one is the value of $\beta$, the lattice coupling.
\bahamas{} uses a \betafile{} where the different values of $\beta$ at which simulations are run can be specified.
This is especially convenient because it avoids to always have to specify the values of $\beta$ from the command line and, moreover, it makes it possible to easily operate on a subset of the simulations just commenting out some lines in the \betafile.
However, it is not just a list of numbers stored somewhere.
The \betafile{} can be thought of as a local configuration file for a set of simulations and it has to be filled out following few simple rules.
\begin{description}[xxxx]
    \item[$\quad\bullet$] Each line corresponds to a different simulation.
    \item[$\quad\bullet$] The value of $\beta$ must appear in first position.
    \item[$\quad\bullet$] All the other fields may be specified in any order using a given prefix.
    \item[$\quad\bullet$] Whole lines or part of them can be commented-out using the \texttt{\#} symbol. 
\end{description}
Using the \betafile{} in a smart way allows to handle simulations in a very convenient and comfortable way.
The possible prefixes that can be used are \texttt{s}, \texttt{i}, \texttt{r}, \texttt{g}, \texttt{t} to specify, respectively, the pseudo-random number generator seed for the simulation, the integrator steps for the molecular dynamics part of the Monte Carlo algorithm, the trajectory number from which the run should be resumed, the goal statistics that should be produced and the time needed on average to make one Monte Carlo step.
The last three are (so far) optional.
However, specifying the \texttt{g} and \texttt{t} fields lets \bahamas{} take care of additional tasks, like determining the walltime of each simulation.

\subsection*{The thermalisation procedure and the $\beta$-folder naming scheme}

In finite temperature lattice QCD, it is common to make a scan in temperature, running a set of simulations at different values of $\beta$.
Since the equilibrium probability distributions of the Markov chains will not be completely different using the same set of physical parameters and varying only $\beta$ by a small amount, it is often convenient (in terms of resources) to perform one single thermalisation from a random configuration at a particular $\beta$ and then re-thermalise at the other $\beta$'s using as starting configuration the last one produced in the first thermalisation.
Later on, the real configuration production can start, using, if needed, several replica (often referred to as \emph{chains} in the following).
Mainly, two patterns are often followed in the thermalisation phase.

\paragraph{Strategy A}

One thermalisation from a \emph{hot} (i.e.\ random) configuration is done at a value of $\beta$ in the $\beta$-range which will be simulated.
Using the arrival configuration as starting one, different chains at different values of $\beta$ are started.
This second thermalisation is referred to as thermalisation from \emph{conf}.
Once finished, the real runs can start and different chains for the same $\beta$ will have completely not-correlated starting points and, then, the data coming from different chains can be safely analysed all together.

\begin{figure}
    \centering
    \includegraphics[width=0.78\textwidth]{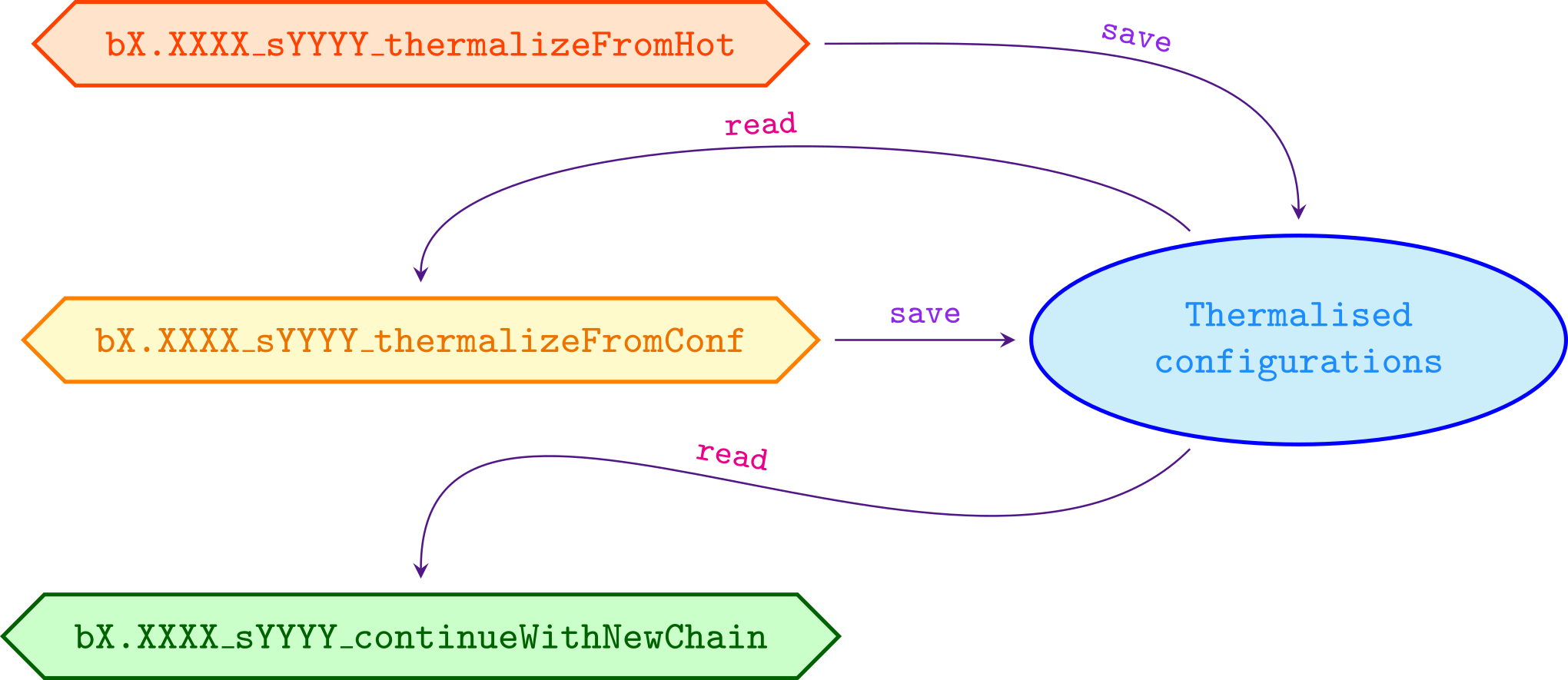}
    \caption{Read and write strategy of \bahamas{} to automatically handle thermalised gauge configurations.}
    \label{fig:readwrite}
\end{figure}%

\paragraph{Strategy B}

Starting new, different chains (at the same $\beta$) from the same starting configuration is the only difference with respect to strategy A.
The advantage is that some resources can be saved, but the first part of the history of the new chains will be highly correlated.

\bigskip

\noindent
Strictly speaking, strategy \emph{A} is the most correct way to proceed, but the \emph{B} alternative is not wrong (it only requires to discard some data in the analysis).
\bahamas{} is able to deal with both strategies and it is up to the user to decide how to behave.
During the job submission or at the end of a simulation in the thermalisation phase, \bahamas{} needs to know only where to store or find the thermalised configurations in order to be able to start the following run(s).
This has been sketched in~\cref{fig:readwrite}, about which any further comment should not be necessary.
Thermalised configurations are stored using a fixed naming scheme, too.
The physical parameters as well as the type of thermalisation are part of the file-name.

\smallskip

All the files referring to the same simulation will be produced in a $\beta$-folder, which is created using a fixed naming scheme.
In general, it is named \texttt{bX.XXXX\_sYYYY\_suffix}, where \texttt{b} and \texttt{s} are the $\beta$ and the seed prefixes, respectively, while the suffix is either \texttt{thermalizeFromHot} or \texttt{thermalizeFromConf} or \texttt{continueWithNewChain}.

\subsection*{The submit-disk and the run-disk}

Often, on clusters, there are several storage systems.
Some are used for permanent files like libraries and executables, some others are just scratch space for temporary files.
What temporary means depends on the supercomputer policy and sometimes files older than a certain amount of time are automatically deleted by the administrators.
The general idea is to store important files on disks for which a backup is done and use the scratch space for anything else.
Unfortunately, sometimes it happens that the backed up space available on the cluster is not large enough and, then, everything is saved on the scratch disk, relying on user backups to avoid accidental losses of data.

\bahamas{} interacts in a natural way with the standard supercomputer structure.
Providing it in the setup procedure with the location of two disks, which we will call \emph{submit-disk} and \emph{run-disk}, it is possible to store the produced files in these two different places.
More precisely, the user will create the typical folder structure on the submit-disk, where the input, standard output and job-script files will be created and from where jobs will be submitted.
At the end of each simulation, the important files (e.g.\ measurements files) will be here copied from the run-disk.
The submit-disk is also the main disk from which \bahamas{} is run.
The run-disk, instead, is used to store the output files and the check-points of a simulation, which can be potentially very large files.

Additional folders needed to use some functionality of \bahamas{}, like the thermalised configurations directory, can be placed anywhere, since it is enough to specify their location in the setup procedure.
Observe that it is possible, and often very convenient, to specify the \textbf{same} location for both disks.
This is not at all a problem and it is actually the way to go whenever, for example, the available space for permanent files is not large enough.

%% file: Examples.tex
\section{\bahamas{} at work}\label{sec:examples}


One of the most interesting and often used features of \bahamas{} is probably the database report functionality.
After having updated the database, an operation which is usually done in background for instance in a screen session, it is possible to get an overview of the full project running \bahamas{} with the \texttt{-{}-database -{}-report} options.
Here below, an example of how its output could look like. 
\begin{center}
    \includegraphics[width=0.88\textwidth]{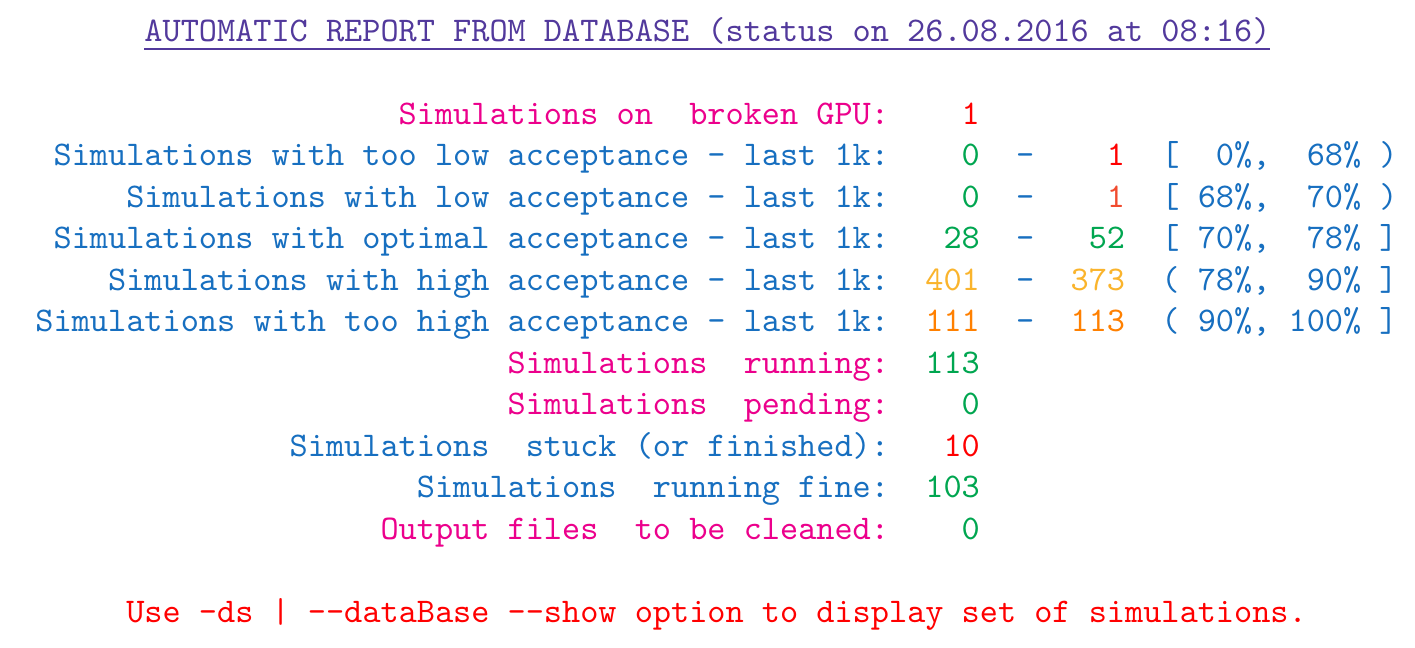}
\end{center}
Even if it refers to a previous version of the software, it clearly shows how simple it is to understand which simulations are problematic.
Moreover, using the \texttt{-{}-database -{}-show} options, it is possible to obtain a list of simulation referring to any line of the report.
This reduces significantly the time required to monitor running simulations.

The \texttt{-{}-database} option of \bahamas{} uses internally the \texttt{liststatus} functionality, which provides the overview of simulations for a given set of the parameters contained in the path.
In both cases, as often in the rest of the code, a coloured output is generated attributing to each colour a precise meaning.
A detailed description of the colour scheme used may be found in the online documentation of the code, while an example of how the output of the \texttt{-{}-liststatus} option appears is reported in \S$3.5$ of~\cite{sciarra}.

%% file: Generalisation.tex
\section{An easy-to-generalise software}\label{sec:generalisation}

\begin{figure}
    \centering
    \includegraphics[width=0.88\textwidth]{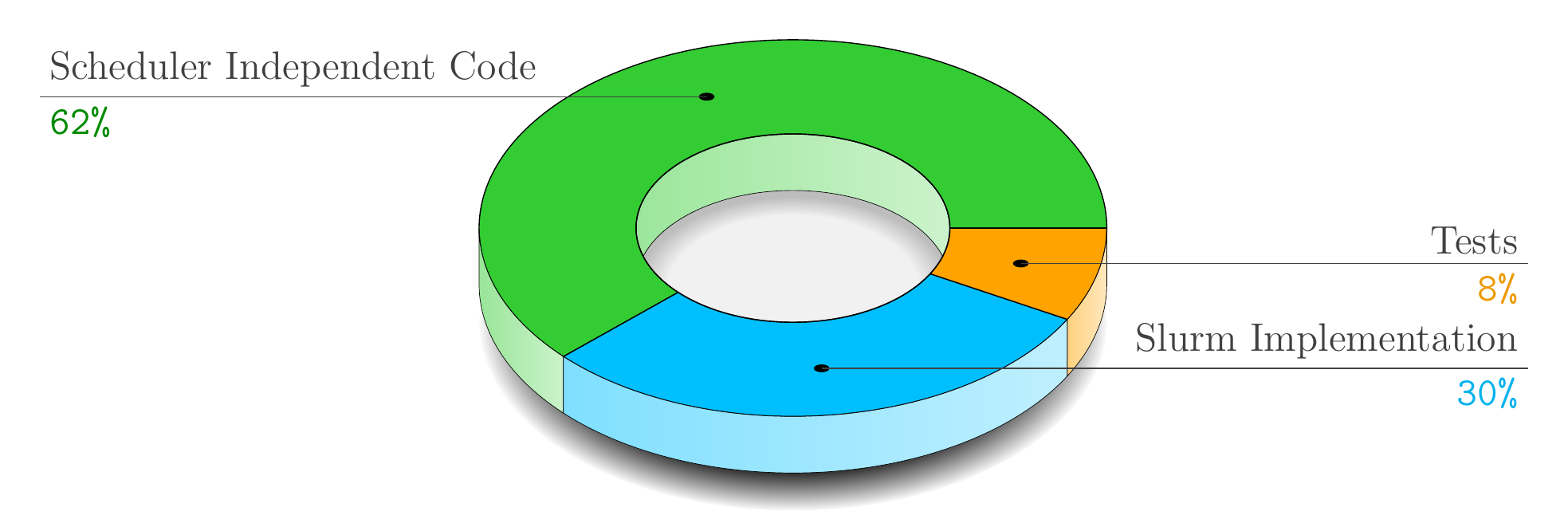}
    \caption{Present structure of the code regarding the dependence of the code on a particular job-scheduler.}
    \label{fig:structure}
\end{figure}

\bahamas{} cannot be considered as the most general tool and, actually, it was born to administrate simulations running the \clqcd~\cite{clqcd} code on clusters provided with the \href{https://slurm.schedmd.com/}{\texttt{slurm}} job-scheduler. 
The decision to release the code led to an intense refactoring phase, in which most of the code was rewritten to easily accommodate generalisations.
At first, operations that do not depend on the job-scheduler were extracted and a sort of interface to use a particular scheduler was created.
In \texttt{bash} it is possible to achieve this in a quite elegant way, using the fact that functions can be called building their name using the content of other variables.
It is then sufficient to include the job-scheduler in the name of the functions, for example as suffix, and build an interface which will call functions in a general way like \texttt{nameOfTheFunction\_\$\{nameOfTheClusterScheduler\}}.
In~\cref{fig:structure} we reported the present structure of the code, where it can be clearly seen that, ignoring tests for a moment, less than half of the software is specific to the scheduler.
We do not exclude to be able to reduce this percentage in the future, extracting further operations that can be considered cluster-independent.   

Another, more complicated aspect, which is still in the design phase, regards how to extract the dependency of \bahamas{} on the production software which it has to run.
Any external code has its own characteristics and there is no standard at all.
Therefore, it is still not clear how to proceed.
For example, the input file, where usually the physical and computational parameters are specified, may have different syntax and different formats, which should be considered.
Some possibilities have been already examined and some work is planned to make \bahamas{} even easier to be extended.

To conclude, some remarks about tests.
Testing a code should be part of the natural code development and there are even frameworks in which tests should exists before the actual code is implemented~\cite{Beck:Test}.
Even though \texttt{bash} is not the ideal programming language to be systematically tested, it does not mean that a \texttt{bash} code \emph{cannot} be tested.
Moreover, refactoring a code without having regression tests (i.e.\ tests that signals if a working software still performs in the same way after some changes) can bring very unpleasant situations.
This is the reason why \bahamas{} was provided with (regression) tests, which constantly help in its development.
Still the situation can and will be improved.
For example, the \href{https://github.com/kward/shunit2}{\texttt{shUnit2}} unit test framework will probably be employed to test single parts of the code. 

%% file: Conclusions.tex
\section{Conclusions and perspectives}\label{sec:conclusions}

\bahamas{} offers a useful tool to automatically administrate and monitor simulations on supercomputers.
Some of its features have been presented and more information is provided in the online documentation.
For the first time, a generic approach to provide a software correctly interacting with different job-schedulers and production codes has been used.
Despite the fact that at the moment, for historical reasons, only \clqcd{} can be used on clusters which make use of the \texttt{slurm} scheduler, the structure of \bahamas{} will naturally accommodate future extensions.
The code is relatively young, but it already shows good potentiality.

Clearly, it is not possible to provide implementations for any existing job-scheduler or for any production code.
In this direction, some work from the user is required and, to facilitate it, some documentation for the developer will be added online in the near future, with particular attention to how to add an implementation for a new scheduler and support for a new production code.  
Ideally, thanks to input by different research groups, \bahamas{} could quickly grow, targeting then a much wider audience.

Concrete plans for the upcoming months target, first of all, some more refactoring improving the software structure and implementation.
This will help to extract the \clqcd{}-dependent parts, which are at the moment hard-coded, and it will hopefully lead to a kind of interface with the production code.
Consequently, it will be clear how to proceed to use a different external program.
At this point we will probably add an implementation for the \href{https://www-03.ibm.com/systems/power/software/loadleveler/}{\texttt{LoadLeveler}} scheduler and, maybe, support for \href{https://github.com/etmc/tmLQCD}{\texttt{tmLQCD}} code.
About upcoming features, we plan to improve and add some information to the simulations overview as well as to the database functionality, which should then make the daily work even easier.

Finally, it is worth commenting on the fact that there is still not an official release of the code, in the sense that nothing like \texttt{v0.1} of \bahamas{} is available.
We would like to spend some more time refactoring the code to then announce a first release, but we are not so far from that.
Soon, we will also start to use the \href{https://github.com/nvie/gitflow}{\texttt{gitflow}} \texttt{git} branching model which will help in having an even more standardised development of \bahamas.